\documentclass[prd,11pt,floatfix,superscriptaddress]{revtex4}

\usepackage{latexsym,amsbsy}
 \usepackage[dvips]{graphicx}
 \DeclareGraphicsExtensions{.eps, .jpg}

\begin{document}
\title{\boldmath
Parameters of cosmological models and recent astronomical observations}



\author{G. S. Sharov} 
\author{E. G. Vorontsova}

\affiliation{Tver state university\\ 170002, Sadovyj per. 35,
Tver, Russia}

\email{german.sharov@mail.ru} 

\begin{abstract}
For different gravitational models we consider
limitations on their parameters coming from recent observational
data for type Ia supernovae, baryon acoustic oscillations, and
from 34 data points for the Hubble parameter $H(z)$ depending on
redshift. We calculate parameters of 3 models describing
accelerated expansion of the universe:  the
$\Lambda$CDM model, the model with generalized  Chaplygin gas
(GCG) and the multidimensional model of I. Pahwa, D.~Choudhury and
T.R.~Seshadri. In particular, for the $\Lambda$CDM model $1\sigma$
estimates of parameters are: $H_0=70.262\pm0.319$
km\,c${}^{-1}$Mpc${}^{-1}$, $\Omega_m=0.276_{-0.008}^{+0.009}$,
$\Omega_\Lambda=0.769\pm0.029$, $\Omega_k=-0.045\pm0.032$. The GCG
model under restriction $\alpha\ge0$ is reduced to the
$\Lambda$CDM model. Predictions of the multidimensional model
essentially depend on 3 data points for $H(z)$ with $z\ge2.3$.
\end{abstract}


\maketitle
\flushbottom

\section{Introduction}\label{Intr}

The most important challenge for cosmologists
 is to explain the accelerated expansion of our
universe that was directly measured for the first time from Type
Ia supernovae observations \cite{Riess98,Perl99}. These supernovae
were used as standard candles, because one can measure their
redshifts $z$ and luminosity distances $D_L$. The observed
dependence $D_L(z)$ based on further measurements
\cite{SNTable,WeinbergAcc12} argues for the accelerated growth of
the cosmological scale factor $a(t)$ at late stage of its
evolution.

This result was confirmed via observations of cosmic microwave
background anisotropy  \cite{WMAP}, baryon acoustic oscillations
(BAO) or large-scale galaxy clustering
\cite{WeinbergAcc12,Eisen05,SDSS}  and other observations
\cite{WeinbergAcc12,WMAP,Plank13}. In particular, our attention
should be paid to measurements of the Hubble parameter $H(z)$ for
different redshifts $z$
\cite{Simon05,Stern10,Moresco12,Blake12,Zhang12,Busca12,Chuang12,Gazta09,Anderson13,Oka13,Delubac14,Font-Ribera13}.
The results of these measurements and estimations are represented
below in Table~\ref{AT1} of Appendix.

The values $H(z)$ were calculated with two methods: evaluation of
the age difference for galaxies with close redshifts in
Refs.~\cite{Simon05,Stern10,Moresco12,Blake12,Zhang12,Busca12,Chuang12}
and the method with BAO analysis
\cite{Gazta09,Anderson13,Oka13,Delubac14,Font-Ribera13}.

In the first method the equality
 \begin{equation}
 a(t)=a_0/(1+z)
 \label{z} \end{equation}
and its consequence
 $$ 
 H(z)=\frac1{a(t)}\frac{da}{dt}=-\frac1{1+z}\frac{dz}{dt}
 $$ 
 are used. Here $a_0\equiv a(t_0)$ is the current value
of the scale factor $a$.

Baryon acoustic oscillations (BAO) are disturbances in the cosmic
microwave angular power spectrum and in the correlation function
of the galaxy distribution, connected with acoustic waves
propagation before the recombination epoch
\cite{WeinbergAcc12,Eisen05}. These waves involved baryons coupled
with photons up to the end of the drag era corresponding to
$z_d\simeq 1059.3$ \cite{Plank13}, when baryons became decoupled
and resulted in a peak in the galaxy-galaxy correlation function
at the comoving sound horizon scale $r_s(z_d)$
\cite{Eisen05,Plank13}.

In Table~\ref{AT2} of Appendix we represent estimations of two
observational manifestations of the BAO effect. These values are
taken from Refs.~\cite{WMAP,BlakeBAO11,Chuang13}, they confirm the
conclusion about accelerated expansion of the universe. In
addition, this data with observations of Type Ia supernovae and
the Hubble parameter $H(z)$ are stringent restrictions on possible
cosmological theories and models.

To explain accelerated expansion of the universe various
cosmological models have been suggested, they include different
forms of dark matter and dark energy in equations of state
 and various modifications of Einstein gravity
\cite{Clifton,Bamba12,CopelandST06}. The most popular among
cosmological models is the $\Lambda$CDM model with a $\Lambda$
term (dark energy) and cold dark matter (see reviews
\cite{Clifton,CopelandST06}). This model with 5\% fraction of
visible baryonic matter nowadays ($\Omega_b=0.05$), 24\% fraction
of dark matter ($\Omega_c=0.24$) and 71\% fraction of dark energy
($\Omega_\Lambda=0.71$) \cite{WMAP} successfully describes
observational data for Type Ia supernovae, anisotropy of cosmic
microwave background, BAO effects and $H(z)$ estimates
 \cite{WeinbergAcc12,WMAP,Plank13}.

 However, there are some problems in the
$\Lambda$CDM model connected with vague nature of dark matter and
dark energy,  with fine tuning of the observed value of $\Lambda$,
which is many orders of magnitude smaller than expected vacuum
energy density, and with surprising proximity $\Omega_\Lambda$ and
$\Omega_m=\Omega_b+\Omega_c$ nowadays, though these parameters
depend on time in different ways (the coincidence problem)
\cite{Clifton,Bamba12,CopelandST06,Kunz12}.

Therefore a large number of alternative cosmological models have
been proposed. They include modified gravity with $f(R)$
Lagrangian \cite{SotiriouF,NojOdinFR}, theories with scalar fields
\cite{CaldwellDS98,KhouryA04}, models with nontrivial equations of
state
\cite{KamenMP01,Bento02,Makler03,LuGX10,LiangXZ11,CamposFP12,XuLu12,LuXuWL11,PaulTh13},
 with extra dimensions
\cite{Mohammedi02,Darabi03,BringmannEG03,PanigrahiZhCh06,MiddleSt11,FarajollahiA10,PahwaChS,GrSh13}
 and many others
\cite{Clifton,Bamba12,CopelandST06,Kunz12}.

Among these gravitational models we concentrate here on the model
with generalized Chaplygin gas (GCG)
\cite{KamenMP01,Bento02,Makler03,LuGX10,LiangXZ11,CamposFP12,XuLu12}.
The equation of state in this model
 \begin{equation}
 p=-B_0/\rho^\alpha
\label{pGCG} \end{equation}
 generalizes the corresponding equation
$p=-B/\rho$ for the original Chaplygin gas model \cite{KamenMP01}.
Generalized Chaplygin gas with EoS (\ref{pGCG}) plays the roles of
both dark matter and dark energy, it is applied to describing
observations of type Ia supernovae, BAO effects,  the Hubble
parameter $H(z)$ and other observational data in various
combinations \cite{Makler03,LuGX10,LiangXZ11,CamposFP12,XuLu12}.

The equation of state similar to Eq.~(\ref{pGCG}) is used in the
multidimensional gravitational model of I. Pahwa, D.~Choudhury and
T.R.~Seshadri \cite{PahwaChS} (the PCS model in references below).
In this model the $1+3+d$ dimensional spacetime 
is symmetric and isotropic in two subspaces: in 3 usual spatial
dimensions and in $d$ additional dimensions. Matter has zero
(dust-like) pressure in usual dimensions and negative pressure
$p_e$ in the form (\ref{pGCG}) in extra dimensions:
\begin{equation}
 T^{\mu}_{\nu} = \mbox{diag}\,(-\rho,0,0,0,p_e,\dots,p_e),
\qquad
 p_e=-B_0{\rho}^{-\alpha}
 \label{Tmn}
\end{equation}
(in Sects.~\ref{Intr}, \ref{Mod} we use units with $c=1$).

In Ref.~\cite{PahwaChS} the important case $d=1$ was omitted. This
case was considered in Ref.~\cite{GrSh13}, where we analyzed
singularities of cosmological solutions in the PCS model
\cite{PahwaChS} and suggested how to modify the equation of state
(\ref{Tmn}) for the sake of avoiding the finite-time future
singularity (``the end of the world'') which is inevitable in the
PCS model. Main advantages of the multidimensional models
\cite{PahwaChS} and \cite{GrSh13} are: naturally arising dynamical
compactification and successful description of the Type Ia
supernovae observations.

In this paper we compare the $\Lambda$CDM model, the model with
generalized Chaplygin gas (GCG) \cite{KamenMP01,Bento02}, and also
the models PCS \cite{PahwaChS} and \cite{GrSh13} with  $d$ extra
dimensions from the point of view of their capacity to describe
recent observational data for type Ia supernovae, BAO and $H(z)$.
In the next section we briefly summarize the dynamics of the
mentioned models, in Sect.~\ref{Observ} we analyze parameters of
the mentioned models resulting in the best description of the
observational data from Ref.~\cite{SNTable} and Appendix.

\section{Models}\label{Mod}

For all cosmological models in this paper the Einstein equations
 \begin{equation}
 G^\mu_\nu=8\pi G T^\mu_\nu+\Lambda\delta^\mu_\nu,
 \label{Eeq}\end{equation}
 determine dynamics of the universe. Here
$T^\mu_\nu$ and $G^\mu_\nu=R^\mu_\nu-\frac12R\delta^\mu_\nu$ are
the energy momentum tensor and the Einstein tensor,  $\Lambda$ is
nonzero only in the $\Lambda$CDM model. The energy momentum tensor
has the form (\ref{Tmn}) in the multidimensional models
\cite{PahwaChS,GrSh13} and the standard form
\begin{equation}
 T^{\mu}_{\nu} = \mbox{diag}\,(-\rho,p,p,p)
 \label{T4}
\end{equation}
in models with $3+1$ dimensions. In the $\Lambda$CDM model
baryonic and dark matter may be considered as one component of
dust-like matter with density $\rho=\rho_b+\rho_{dm}$, so we
suppose $p=0$ in Eq.~(\ref{T4}). The fraction of relativistic
matter (radiation and neutrinos) is close to zero for observable
values $z\le2.3$. In the GCG model
\cite{KamenMP01,Bento02,Makler03,LuGX10,LiangXZ11,CamposFP12,XuLu12}
pressure $p$ in the form (\ref{pGCG}) plays the role of dark
energy, corresponding to the $\Lambda$ term in the $\Lambda$CDM
model.

For the Robertson-Walker metric with the curvature sign $k$
 \begin{equation}
 ds^2 = -dt^2+a^2(t)\Big[(1-k r^2)^{-1}dr^2+r^2
 d\Omega\Big]
 \label{metrRW}
 \end{equation}
 the Einstein equations (\ref{Eeq}) are reduced to the system
\begin{eqnarray}
3\frac{\dot{a}^2+k}{a^2}=8\pi G\rho+\Lambda,\label{Esysa}\\
 \dot{\rho}=-3\frac{\dot{a}}{a}(\rho+p).\label{Esys2}
\end{eqnarray}
 Eq.~(\ref{Esys2}) results from the continuity
condition $T^{\mu}_{\nu; \mu}=0$, the dot denotes the time
derivative.

 Using the present time values  of the Hubble constant
 and the critical density
   \begin{equation}
 H_0=\frac{\dot a} a\Big|_{t=t_0}=H\Big|_{z=0},\qquad
 \rho_{cr}=\frac{3H_0^2}{8\pi G},
 \label{rocr}\end{equation}
 we introduce dimensionless time $\tau$, densities
$\bar{\rho}_i$, pressure $\bar{p}$ and logarithm of the scale
factor \cite{PahwaChS,GrSh13}:
\begin{equation}
\tau=H_0t,\qquad\bar{\rho}=\frac{\rho}{\rho_{cr}},\qquad
\bar{\rho}_b=\frac{\rho_b}{\rho_{cr}},\qquad
\bar{p}=\frac{p}{\rho_{cr}},\qquad {\cal A}=\log\frac a{a_0}.
 \label{tau} \end{equation}
 We denote derivatives with
respect to $\tau$ as primes and rewrite the system (\ref{Esysa}),
(\ref{Esys2}) 
 \begin{eqnarray}
{\cal A}'(\tau)&=&\sqrt{\bar{\rho}+\Omega_\Lambda+\Omega_ke^{-2{\cal A}}}, \label{Asy} \\
\bar{\rho}'(\tau)&=&-3{\cal A}'(\bar{\rho}+\bar{p}). \label{rhsy}
 \end{eqnarray}
 Here
 \begin{equation}
\Omega_m=\frac{\rho(t_0)}{\rho_{cr}},\qquad
\Omega_\Lambda=\frac{\Lambda}{3H_0^2},\qquad
 \Omega_k=-\frac{k}{a_0^2H_0^2}
 \label{Omega1} \end{equation}
 are present time fractions of matter
($\Omega_m=\Omega_b+\Omega_c$), dark energy and curvature in the
equality
 \begin{equation}
\Omega_m+\Omega_\Lambda+ \Omega_k=1,
 \label{sumOm} \end{equation}
resulting from Eq.~(\ref{Esysa}) if we fix $t=t_0$.

If we know an equation of state $\bar{p}=\bar{p}(\bar{\rho})$ for
 any model, we can solve the Cauchy problem for the system (\ref{Asy}),
(\ref{rhsy}) including initial conditions for variables
(\ref{tau}) at the present epoch $t=t_0$ (here and below $t=t_0$
corresponds to $\tau=1$)
 \begin{equation}
 {\cal A}\big|_{\tau=1}=0,\qquad
 \bar{\rho}\big|_{\tau=1}=\Omega_m.
 \label{init1} \end{equation}

In the $\Lambda$CDM model Eq.~(\ref{rhsy}) yields
$\bar{\rho}=\Omega_m e^{-3{\cal A}}=\Omega_m(1+z)^3$, so we solve
only equation (\ref{Asy})
 \begin{equation}
 {{\cal A}'}^2=\frac{H^2}{H_0^2}=\Omega_m e^{-3{\cal A}}+
 \Omega_\Lambda+\Omega_ke^{-2{\cal A}}.
  \label{ALCDM} \end{equation}
 with the first initial condition
(\ref{init1}).

Equation (\ref{rhsy}) may be solved also and in the GCG model, but
in this case we are to decompose all matter into two components
\cite{LuGX10,LiangXZ11,CamposFP12,XuLu12,LuXuWL11}. One of these
components is usual dust-like matter including baryonic matter;
the other component is generalized Chaplygin gas with density
$\rho_g\equiv\rho_{GCG}$ (and corresponding
$\bar{\rho}_g=\rho_g/\rho_{cr}$). If the first component is pure
baryonic and the latter describes both dark matter and dark
energy, equations of state are:
 \begin{equation}
\bar{\rho}=\bar{\rho}_b+\bar{\rho}_g,\qquad\bar{p}_b=0,\qquad\bar{p}=\bar{p}_g=-B\,(\bar{\rho_g})^{-\alpha}
 \label{EoSC}
 \end{equation}

If we use the integrals $\bar{\rho}_b=\Omega_b e^{-3{\cal A}}$ and
$\bar{\rho}_g=\big[B+Ce^{-3{\cal
A}(1+\alpha)}\big]^{1/(1+\alpha)}$ of  Eq.~(\ref{rhsy}) for these
components, equation (\ref{Asy}) takes the form
 \cite{Makler03,LuGX10,LiangXZ11,CamposFP12,XuLu12,LuXuWL11}
 \begin{equation}
 {{\cal A}'}^2=\frac{H^2}{H_0^2}=\Omega_b e^{-3{\cal A}}+
 (1-\Omega_b-\Omega_k)\Big[B_s+(1-B_s)\,e^{-3{\cal
A}(1+\alpha)}\Big]^{1/(1+\alpha)}+\Omega_ke^{-2{\cal A}}.
  \label{AGCG} \end{equation}
We solve this equation  with the initial condition (\ref{init1}) $
{\cal A}\big|_{\tau=1}=0$. The dimensionless constant $B_s$
\cite{XuLu12,LuXuWL11} (it is denoted $A_s$ in
Refs.~\cite{LuGX10,LiangXZ11}) is expressed via $B$ or $B_0$:
  \begin{equation}
B_s=B\cdot(1-\Omega_b-\Omega_k)^{-1-\alpha},\qquad
B=B_0\,\rho_{cr}^{-1-\alpha}.
  \label{Bs} \end{equation}

For the multidimensional model PCS \cite{PahwaChS} and the model
\cite{GrSh13} in spacetime with $1+3+d$ dimensions the following
metric is used \cite{PahwaChS}:
 \begin{equation}
 ds^2 = -dt^2+a^2(t)\left(\frac{dr^2}{1-k r^2} + r^2
 d\Omega\right)+b^2(t)\left(\frac{d R^2}{1-k_2 R^2}+R^2 d\Omega_{d-1}
\right).
 \label{metricInd}
 \end{equation}
  Here  $b(t)$ and $k_2$  are the scale factor and
curvature sign in extra dimensions (along with $a$ and $k$ for
usual dimensions). For cosmological solutions in
Refs.~\cite{PahwaChS,GrSh13}
 the scale factor $a(t)$ grows while $b(t)$ diminishes,
in other words, some form of dynamical compactification
\cite{Mohammedi02,Darabi03,BringmannEG03,PanigrahiZhCh06,MiddleSt11,FarajollahiA10,PahwaChS}
takes place, a size of compactified $b$ is small enough to play no
essential role at the TeV scale.

In Refs.~\cite{PahwaChS,GrSh13} the authors considered only one
component of their matter. Here we generalize these models and
introduce the ``usual'' component with density $\bar{\rho}_b$ and
the ``exotic'' component with $\bar{\rho}_e=\rho_e/\rho_{cr}$ and
pressure $\bar{p}_e=p_e/\rho_{cr}$ in extra dimensions similarly
to Eq.~(\ref{EoSC}):
 \begin{equation}
\bar{\rho}=\bar{\rho}_b+\bar{\rho}_e,\qquad\bar{p}_e=-B\,(\bar{\rho_e})^{-\alpha}
 \label{EoPCS} \end{equation}

 Dynamical equations for the models \cite{PahwaChS,GrSh13} result
from the Einstein equations (\ref{T4}) with $\Lambda=0$ and the
energy momentum tensor (\ref{Tmn}), (\ref{EoPCS}). In our notation
(\ref{tau}) with  ${\cal B}=\log\big(b/b_0\big)$ (where
$b_0=b(t_0)$) these equations for $k_2=0$ and $d>1$ are
\cite{PahwaChS,GrSh13}
 \begin{eqnarray}
{\cal A}''&=&\frac1{d+2}\Big[ d(d-1)\,{\cal B}'\big(\frac12{\cal
B}'-{\cal A}'\big)-3(d+1)\,{{\cal A}'}^2 -3d\bar{p}_e
 +(2d+1)\Omega_ke^{-2{\cal A}}\Big], \quad\label{Ad} \\
\bar{\rho}_b'&=&-\bar{\rho}_b(3{\cal A}'+d{\cal B}'),\qquad
\bar{\rho}_e'=-3\bar{\rho_e}{\cal
A}'-d(\bar{\rho_e}+\bar{p}_e)\,{\cal B}',
\label{rhod}\\
{\cal B}'&=&(d-1)^{-1}\Big[-3{\cal A}'+\sqrt{3\big[(d+2)\,{{\cal
A}'}^2+2(d-1)\, (\bar{\rho}+\Omega_ke^{-2{\cal A}})\big]/d}\Big].
\label{Bd}
 \end{eqnarray}
 If $d=1$ one should use
\cite{GrSh13}
 \begin{equation}
 {\cal B}'=(\bar{\rho}+\Omega_ke^{-2{\cal A}})/{\cal A}'-{\cal A}'
 \label{B1} \end{equation}
 instead of Eq.~(\ref{Bd}).

For the system (\ref{Ad})~--~(\ref{rhod}) the initial conditions
include Eqs.~(\ref{init1}) and the additional condition
 \begin{equation}
 {\cal A}'\big|_{\tau=1}=1
 \label{init2} \end{equation}
resulting from definitions of ${\cal A}$ (\ref{tau}) and $H_0$
(\ref{rocr}):
 $${\cal A}'(\tau)=\frac d{d\tau}\log\frac a{a_0}=\frac1{H_0}
\frac{\dot a}a.$$

For the model PCS \cite{PahwaChS,GrSh13} we have the analog of
Eq.~(\ref{sumOm})
 \begin{equation}
\Omega_m+\Omega_B+ \Omega_k=1,
 \label{sumOmI} \end{equation}
 resulting from  Eqs.~(\ref{Bd}) or (\ref{B1}) at $\tau=1$. Here
$\Omega_B=-d\big(B'+\frac{d-1}6{B'}^2\big)\big|_{\tau=1}$ is the
contribution from $d$ extra dimensions.

The models $\Lambda$CDM, GCG, PCS with suitable values of model
parameters have cosmological solutions describing accelerated
expansion of the universe
\cite{WMAP,Plank13,Makler03,LuGX10,LiangXZ11,CamposFP12,XuLu12,PahwaChS,GrSh13}.
We consider restrictions on these parameters coming from recent
observational data for type Ia supernovae \cite{SNTable}, BAO
\cite{WMAP,BlakeBAO11,Chuang13} and from measuring the Hubble
parameter $H(z)$
\cite{Simon05,Stern10,Moresco12,Blake12,Zhang12,Busca12,Chuang12,Gazta09,Anderson13,Oka13,Delubac14,Font-Ribera13},
 (Tables~\ref{AT2}, \ref{AT1}).

\section{Observational data and model parameters }\label{Observ}

Recent observational data on Type Ia supernovae in the Union2.1
compilation \cite{SNTable} include redshifts $z=z_i$ and distance
moduli $\mu_i$ with errors $\sigma_i$ for $N_S=580$ supernovae.
The distance modulus
$\mu_i=\mu(D_L)=5\log\big(D_L/10\mbox{pc}\big)$ is logarithm of
the luminosity distance \cite{Plank13,Clifton}:
 \begin{equation}
 D_L(z)=\frac{c\,(1+z)}{H_0\sqrt{|\Omega_k|}}\mbox{\,Sin}_k
 \bigg(H_0\sqrt{|\Omega_k|}\int\limits_0^z\frac{d\tilde z}{H(\tilde
 z)}\bigg),\quad\;
\mbox{Sin}_k(x)=\left\{\begin{array}{ll} \sinh x, &\Omega_k>0,\\
 x, & \Omega_k=0,\\ \sin x, & \Omega_k<0. \end{array}\right.
  \label{DL} \end{equation}
 In particular, for the flat universe ($k=\Omega_k=0$) the expression (\ref{DL})
is
 $$
 D_L=c\,(1+z)\int\limits_0^z\frac{d\tilde z}{H(\tilde z)}
 =\frac{ca_0^2}{H_0a(\tau)}\int\limits_\tau^1\frac{d\tilde\tau}{a(\tilde\tau)},
 $$

To describe the Type Ia supernovae data \cite{SNTable} we fix
values of model parameters $p_1,p_2,\dots$ for the chosen model
$\Lambda$CDM, GCG or PCS and calculate dependence of the scale
factor $a(\tau)$ on dimensionless time $\tau$. Further, we
calculate numerically the integral expression (\ref{DL}) and the
distance modulus $\mu(\tau)$. For each value of redshift $z_i$ in
the table \cite{SNTable} we find the corresponding $\tau=\tau_i$
with using linear approximation in Eq.~(\ref{z}) and the
theoretical value $\mu_{th}=\mu(\tau_i,p_1,p_2,\dots)$ from the
dependence $\mu(\tau)$ (\ref{DL}).

We search a good fit between theoretical predictions $\mu_{th}$
and the observed data  $\mu_i$ as the minimum of
 \begin{equation}
 \chi^2_S(p_1,p_2,\dots)=\sum_{i=1}^{N_S}
 \frac{\big[\mu_i-\mu_{th}(z_i,p_1,p_2,\dots)\big]^2}{\sigma_i^2}
  \label{chiS} \end{equation}
or the maximum of the corresponding likelihood function
 ${\cal L}_S(p_1,p_2,\dots)=\exp(-\chi^2_S/2)$
in the space of model parameters $p_1,p_2,\dots$

\begin{figure}[th]
  \centerline{\includegraphics[scale=0.76,trim=3mm 0mm 2mm 6mm]{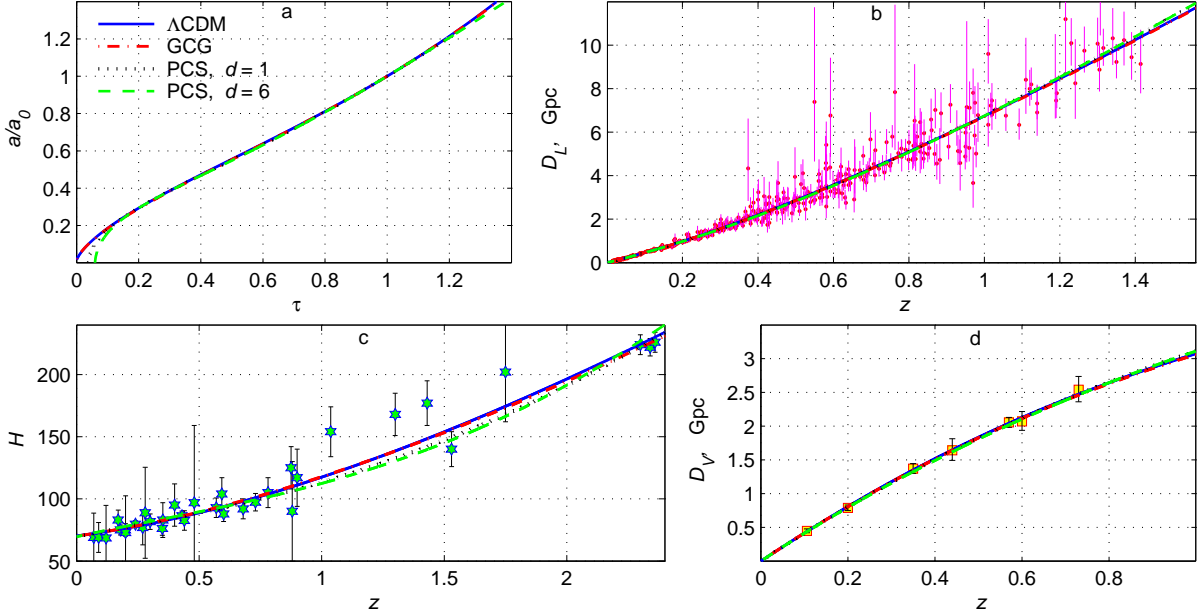}}
  \caption{\small For the models $\Lambda$CDM, GCG, PCS ($d=1$
and $d=6$) with the optimal values of model parameters from
Table~\ref{Optim} we present (a) the scale factor $a(\tau)$; (b)
the luminosity distance $D_L(z)$  and the Type Ia supernovae data
\cite{SNTable}; (c) dependence $H(z)$ with the data points from
Table~\ref{AT1} and (d) the distance (\ref{DV}) $D_V(z)$ with the
data points from Table~\ref{AT2}.}
  \label{F1}
\end{figure}

The Type Ia supernovae data \cite{SNTable} and the best fits for
the mentioned models $\Lambda$CDM, GCG and PCS are shown in
Fig.~\ref{F1}b in $z,D_L$ plane. Details of the optimization
procedure are described below.

Model predictions for the Hubble parameter $H(z)=\dot a/a=H_0{\cal
A}'(\tau)$ we compare with observational data
\cite{Simon05,Stern10,Moresco12,Blake12,Zhang12,Busca12,Chuang12,Gazta09,Anderson13,Oka13,Delubac14,Font-Ribera13},
from Table~\ref{AT1} (Fig.~\ref{F1}c) and use the $\chi^2$
function similar to (\ref{chiS}):
\begin{equation}
 \chi^2_H(p_1,p_2,\dots)=\sum_{i=1}^{N_H}
 \frac{\big[H_i-H_{th}(z_i,p_1,p_2,\dots)\big]^2}{\sigma_{H,i}^2}.
  \label{chiH} \end{equation}
 Here $N_H=34$, theoretical values $H_{th}(z_i,\dots)=H_0{\cal A}'\big(\tau(z_i)\big)$
are obtained from the calculated dependence ${\cal A}(\tau)$ and the equality
(\ref{z}) $z=e^{-{\cal A}}-1$.

The observational data for BAO \cite{WMAP,BlakeBAO11,Chuang13}
(Table~\ref{AT2}) includes two measured values \cite{Eisen05}
 \begin{equation}
 d_z(z)= \frac{r_s(z_d)}{D_V(z)}
  \label{dz} \end{equation}
 and
 \begin{equation}
  A(z) = \frac{H_0\sqrt{\Omega_m}}{cz}D_V(z).
  \label{Az} \end{equation}
 They are connected with the distance \cite{WMAP,Eisen05,Plank13}
 \begin{equation}
D_V(z)=\bigg[\frac{cz D_L^2(z)}{(1+z)^2H(z)}\bigg]^{1/3},
 \label{DV} \end{equation}
 expressed here via the luminosity distance (\ref{DL}).

The BAO observations \cite{WMAP,BlakeBAO11,Chuang13} in
Table~\ref{AT2} are not independent. So the $\chi^2$ function for
the values (\ref{dz}) and (\ref{Az})
 \begin{equation}
 \chi^2_B(p_1,p_2,\dots)=(\Delta d)^TC_d^{-1}\Delta d+
(\Delta { A})^TC_A^{-1}\Delta { A}.
  \label{chiB} \end{equation}
 includes the columns 
$\Delta d=[d_{z,th}(z_i,p_1,\dots)-d_z(z_i)]$, $\Delta
A=[A_{th}(z_i,p_1,p_2,\dots)-{ A}(z_i)]$, $i=1,\dots,N_B$ and the
covariance matrices $C_d^{-1}$ and $C_A^{-1}$
\cite{WMAP,BlakeBAO11} described in Appendix.

\smallskip

The best fits to the observational data for Type Ia supernovae
\cite{SNTable}, $H(z)$ and BAO data from Tables~\ref{AT2},
\ref{AT1} are presented in Fig.~\ref{F1} for the models
$\Lambda$CDM, GCG and PCS (with $d=1$ and $d=6$). The values of
model parameters are tabulated below in Table~\ref{Optim}. They
are optimal from the standpoint of minimizing the sum of all
$\chi^2$  (\ref{chiS}), (\ref{chiH}) and (\ref{chiB}):
 \begin{equation}
\chi^2_\Sigma=\chi^2_S+\chi^2_H+\chi^2_B.
 \label{chisum} \end{equation}

Predictions of different models in Fig.~\ref{F1} are rather close,
in particular, the curves for the models $\Lambda$CDM and GCG
practically coincide. The Hubble parameter $H(z)$ in
Fig.~\ref{F1}c is measured in km\,c${}^{-1}$Mpc${}^{-1}$, the
distances $D_L(z)$ and $D_V(z)$ in Fig.~\ref{F1}b,\,d are in Gpc.

The data points for $D_V(z)=r_s(z_d)/d_z(z)$ in Fig.~\ref{F1}d are
calculated from $d_z(z_i)$ in Table~\ref{AT2}. Here the error
boxes include the data spread between the recent estimations of
the comoving sound horizon size:
  \begin{equation}
 r_s(z_d)=147.49\pm 0.59\mbox{ Mpc \cite{Plank13}},\qquad
 r_s(z_d)=153.3\pm 2.0\mbox{ Mpc \cite{Anderson13,BlakeBAO11}}.
  \label{rs} \end{equation}

\subsection{$\Lambda$CDM model}

In the $\Lambda$CDM model we use three free parameters $H_0$,
$\Omega_m$ and $\Omega_\Lambda$ in Eq.~(\ref{ALCDM}) for
describing the considered observational data at $z\le2.3$. For the
Hubble constant  $H_0$ different approaches result in different
estimations. In particular, observations of Cepheid variables in
the project Hubble Space Telescope (HST) give the recent estimate
$H_0=73.8\pm2.4$ km\,c${}^{-1}$Mpc${}^{-1}$ \cite{Riess11}. On the
other hand, the satellite projects Planck Collaboration (Planck)
\cite{Plank13} and Wilkinson Microwave Anisotropy Probe (WMAP)
\cite{WMAP} for observations of cosmic microwave background
anisotropy result in the following values (in
km\,c${}^{-1}$Mpc${}^{-1}$):
 \begin{equation}\begin{array}{ll}
 H_0=67.3 \pm 1.2 & \mbox{ (Planck \cite{Plank13})},\\
 H_0=69.7\pm2.4 & \mbox{ (WMAP \cite{WMAP})},\\
 H_0=73.8\pm2.4  &  \mbox{ (HST \cite{Riess11})}.
 \end{array}
 \label{H0}\end{equation}
 The nine-year results from WMAP \cite{WMAP} include also the
estimate $H_0=69.33\pm0.88$ km\,c${}^{-1}$Mpc${}^{-1}$ with added
recent BAO and $H_0$ observations.

For the $\Lambda$CDM model many authors
\cite{WMAP,Plank13,Tonry03,Knop03,Kowalski08,ShiHL12,FarooqMR13,FarooqR13,Farooqth}
calculated the best fits for parameters $H_0$, $\Omega_m$ and
$\Omega_\Lambda$ for describing the Type Ia supernovae, $H(z)$ and
BAO data in various combinations. In
Refs.~\cite{ShiHL12,FarooqMR13,FarooqR13,Farooqth} some other
cosmological models were compared with the $\Lambda$CDM model. In
particular, the authors \cite{ShiHL12} compared 8 models with two
information criteria including minimal $\chi^2$ and the number of
model parameters. Optimal values of these parameters were pointed
out in Ref.~\cite{ShiHL12} with the exception of $H_0$, though
$H_0$ is the important parameter for all 8 models.

In Refs.~\cite{FarooqMR13,FarooqR13,Farooqth} the $\Lambda$CDM,
XCDM and $\phi$CDM models were applied to describe the supernovae,
$H(z)$ and BAO data. For all mentioned models the authors
\cite{FarooqMR13,FarooqR13,Farooqth} fixed two values of the
Hubble constant $H_0=68\pm2.8$ \cite{GottV01} and $H_0=73.8\pm2.4$
km\,c${}^{-1}$Mpc${}^{-1}$ \cite{Riess11} and searched optimal
values of other model parameters. But they did not estimated the
best choice of $H_0$ among these two values and in the segment
between them.

In this paper we pay the special attention to dependence of
$\chi^2_\Sigma$  minima on $H_0$. This dependence is very
important if we compare different cosmological models.

The results of calculations
\cite{WMAP,Plank13,Kowalski08,ShiHL12,FarooqMR13,FarooqR13,Farooqth},
 as usual, are presented as level lines for
the functions $\chi^2(p_1,p_2)$ or ${\cal
L}_S(p_1,p_2)=\exp(-\chi^2_S/2)$ of two parameters at $1\sigma$
(68.27\%), $2\sigma$ (95.45\%) and $3\sigma$ (99.73\%) confidence
levels. In particular, if a value $H_0$ is fixed, these two
parameters for the $\Lambda$CDM model may be $\Omega_m$ and
$\Omega_\Lambda$.

\begin{figure}[th]
  \centerline{\includegraphics[scale=0.79,trim=5mm 0mm 5mm 6mm]{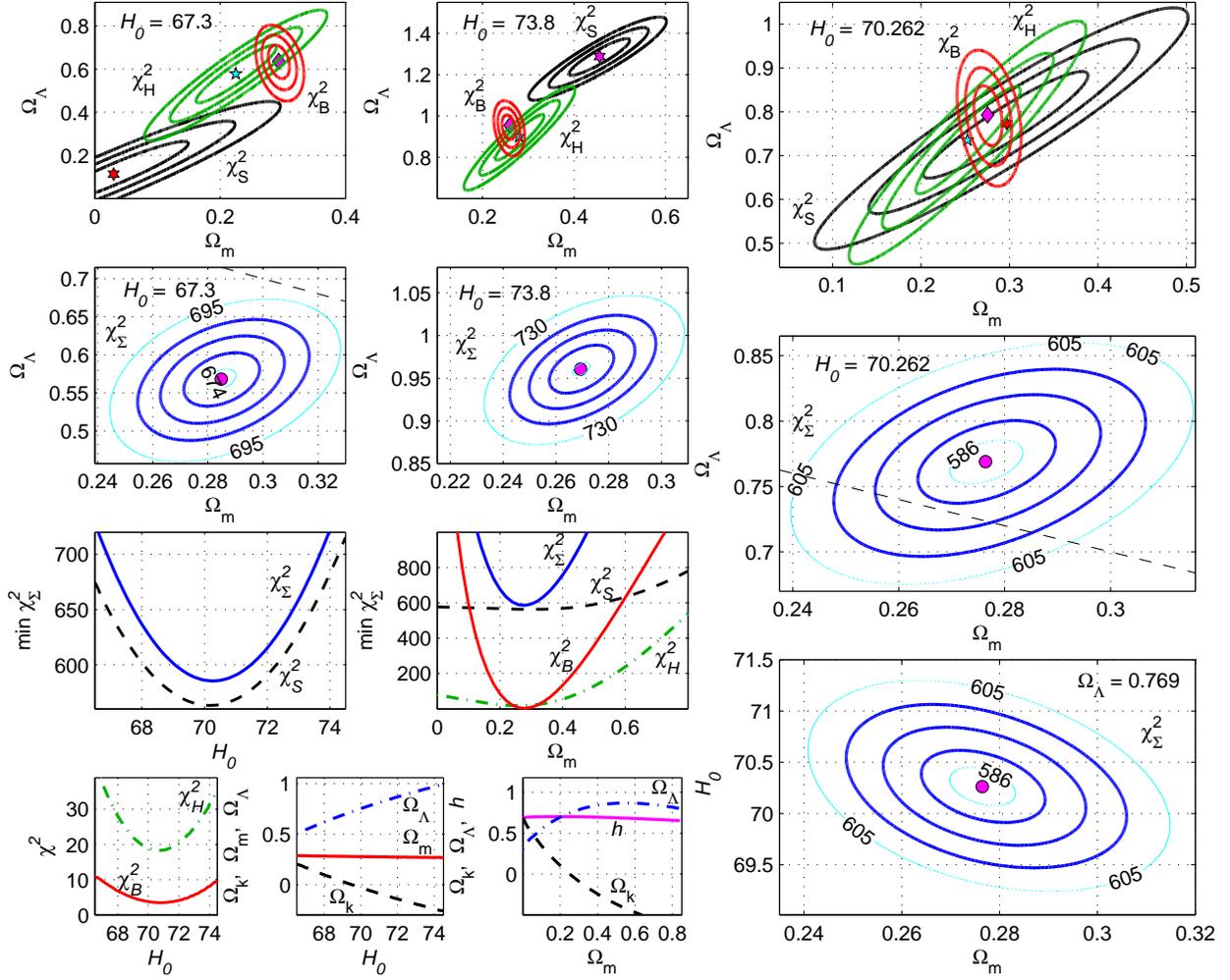}}
  \caption{\small The $\Lambda$CDM model. For the values $H_0$ (\ref{H0})
and the optimal value $H_0=70.26$ km\,c${}^{-1}$Mpc${}^{-1}$ level
lines are drawn at $1\sigma$, $2\sigma$ and $3\sigma$ (thick
solid) for $\chi^2_S(\Omega_m,\Omega_\Lambda)$ (black), for
$\chi^2_H(\Omega_m,\Omega_\Lambda)$ (green) and
$\chi^2_B(\Omega_m,\Omega_\Lambda)$ (red in the top  row), the sum
(\ref{chisum}) $\chi^2_\Sigma(\Omega_m,\Omega_\Lambda)$ (the
middle row), $\chi^2_\Sigma(\Omega_m,H_0)$ for
$\Omega_\Lambda=0.758$ (the bottom-right panel); dependence of
$\min\chi^2_\Sigma$, its fractions $\chi^2$ and parameters of a
minimum point on $H_0$ and on $\Omega_m$. }
  \label{F2}
\end{figure}

In Fig.~\ref{F2} we use this scheme for 3 fixed values $H_0$
(\ref{H0}) indicated on the panels (including  the optimal value
$H_0=70.262$ km\,c${}^{-1}$Mpc${}^{-1}$) and draw level lines of
the functions (\ref{chiS}), (\ref{chiH}), (\ref{chiB}) and
(\ref{chisum}) $\chi^2(\Omega_m,\Omega_\Lambda)$ in the
$\Omega_m,\Omega_\Lambda$ plane and for
$\chi^2_\Sigma(\Omega_m,H_0)$ with fixed $\Omega_\Lambda=0.769$ in
the bottom-right panel. The points of minima are marked in
Fig.~\ref{F2} as hexagrams for $\chi^2_S$, pentagrams for
$\chi^2_H$, diamonds for $\chi^2_B$ and circles for
$\chi^2_\Sigma$. Minimal values of the functions $\chi^2$
(\ref{chiS}), (\ref{chiH}), (\ref{chiB}) and (\ref{chisum}) at
these points are tabulated in Table~\ref{TLCDM} so we can compare
efficiency of this description for different $H_0$. For the same
purpose we point out the corresponding values $\chi^2$ for some
level lines in Fig.~\ref{F2} and present the dependence of minima
$\min\chi^2_\Sigma$ on $H_0$ and on $\Omega_m$ in the left bottom
panels of Fig.~\ref{F2}. Here  we denote
$\min\chi^2_\Sigma(H_0)=\min\limits_{\Omega_m,\Omega_\Lambda}\chi^2_\Sigma$,
$\min\chi^2_\Sigma(\Omega_m)=\min\limits_{H_0,\Omega_\Lambda}\chi^2_\Sigma$
and graphs of the fractions $\chi^2_S$,  $\chi^2_H$, $\chi^2_B$ in
$\min\chi^2_\Sigma(H_0)$ are also shown.

In the bottom panels we present how parameters of a minimum point
of $\chi^2_\Sigma$ depend on $H_0$ and on $\Omega_m$. In
particular, for the  dependence on $H_0$ the coordinates
$\Omega_m(H_0)$ and $\Omega_\Lambda(H_0)$ of this point are
calculated, the value $\Omega_k$ is determined from
Eq.~(\ref{sumOm}). For the dependence on $\Omega_m$ we also
present the graph $h(\Omega_m)$, where $h=H_0/100$.

We see in  Fig.~\ref{F2} and in Table~\ref{TLCDM} that the
dependence of  $\min\chi^2_\Sigma(H_0)$ 
is  appreciable and significant. This function has the distinct
minimum and achieves its minimal value $585.35$ at
$H_0\simeq70.26$. The optimal values of the $\Lambda$CDM model
parameters $\Omega_m\simeq0.276$, $\Omega_\Lambda\simeq0.769$,
corresponding to this minimum are presented in Table~\ref{Optim},
these values are taken for the $\Lambda$CDM curves in
Fig.~\ref{F1}.

The mentioned sharp dependence of $\min\chi^2_\Sigma$ on $H_0$ is
connected with two factors: (1) the similar dependence of the main
contribution $\chi^2_S(H_0)$ shown in the same panel; (2) the
large shift of the minimum point for $\chi^2_S$ in the
$\Omega_m,\Omega_\Lambda$ plane corresponding to $H_0$ growth. For
$H_0=68$ and $73.8$ km\,c${}^{-1}$Mpc${}^{-1}$ this  minimum point
is far from the similar points of $\chi^2_H$ and $\chi^2_B$. Only
for $H_0$ close to 70 km\,c${}^{-1}$Mpc${}^{-1}$ all these three
minimum points are near each other (the top-right panel in
Fig.~\ref{F2}).

Only the value $H_0= 69.7$ km\,c${}^{-1}$Mpc${}^{-1}$ in
Table~\ref{TLCDM} is close to the optimal value  in
Table~\ref{Optim}. We may conclude that the values of the Hubble
constant $H_0=68$ and $73.8$ km\,c${}^{-1}$Mpc${}^{-1}$ taken in
Refs.~\cite{FarooqMR13,FarooqR13,Farooqth}, unfortunately, lie to
the left and to the right from the optimal value $H_0\simeq70$
km\,c${}^{-1}$Mpc${}^{-1}$. We see the significant difference
between the large values $\min\chi^2_\Sigma=673.64$ or $707.84$
for the too small and too large values of $H_0$ in
Table~\ref{TLCDM} and the optimal value $\min\chi^2_\Sigma=585.35$
for $H_0=70.262$ in Table~\ref{Optim}.

In the middle row panels of Fig.~\ref{F2} with $\chi^2_\Sigma$ the
flatness line $\Omega_m+\Omega_\Lambda=1$ (or $\Omega_k=0$) is
shown as the black dashed straight line. This line shows that only
for $H_0$ close to the optimal value from Table~\ref{Optim} the
following recent observational limitations on the $\Lambda$CDM
model parameters (\ref{Omega1}) from surveys \cite{WMAP,Plank13}
 \begin{equation}
 \begin{array}{llll}
& \Omega_m= 0.279\pm 0.025, & & \Omega_m= 0.314\pm 0.02\\
 \mbox{WMAP \cite{WMAP}: \ }&\Omega_\Lambda= 0.721\pm 0.025,\quad &
 \mbox{ \ \ Planck \cite{Plank13}: \ }&\Omega_\Lambda= 0.686\pm 0.025,\\
 & \Omega_k=-0.0027^{+0.0039}_{-0.0038}; & &
 \Omega_k=-0.0005^{+0.0065}_{-0.0066}\end{array}
 \label{Omk} \end{equation}
 are satisfied on $1\sigma$ or $2\sigma$ level.
For $H_0=67.3$ and $73.8$ km\,c${}^{-1}$Mpc${}^{-1}$ the optimal
values of parameters $\Omega_m$, $\Omega_\Lambda$, $\Omega_k$ in
Table~\ref{TLCDM} are far from restrictions (\ref{Omk}) for
$\Omega_k$ even on $3\sigma$ level.

\begin{table}[ht]
\caption{ The $\Lambda$CDM model. For given $H_0$ (\ref{H0}) the
calculated minima of $\chi^2_S$, $\chi^2_H$, $\chi^2_B$ and
$\chi^2_\Sigma$ with  $\Omega_m$, $\Omega_\Lambda$, $\Omega_k$
correspondent to $\min\chi^2_\Sigma$.}
\begin{center}
\begin{tabular}{||c||c|c|c||c|c|c|c||}  \hline
 $H_0$  &  $\min\chi^2_S$ &  $\min\chi^2_H$ & $\min\chi^2_B$
 &  $\min\chi^2_\Sigma$ &$\Omega_m$ & $\Omega_\Lambda$&$\Omega_k$\\ \hline
 67.3& 599.37 & 18.492& 5.548& 673.64 & 0.285& 0.568& 0.147\\ \hline
 69.7& 562.73 & 17.993& 3.517& 588.53 & 0.278& 0.734&$-0.012$\\ \hline
 73.8& 639.90 & 19.466& 5.322& 707.84 & 0.269& 0.961&$-0.230$\\ \hline
 \end{tabular}
\end{center}
 \label{TLCDM}\end{table}

Graphs of the optimal values  $\Omega_m$, $\Omega_\Lambda$ and
$\Omega_k$ depending on $H_0$ are presented in the second bottom
panel. We see that the  value  $\Omega_m$ weakly  depends on
$H_0$, but $\Omega_\Lambda$ and $\Omega_k$ satisfy conditions
(\ref{Omk}) only for $H_0\simeq70$ km\,c${}^{-1}$Mpc${}^{-1}$.

The dependence of $\min\limits_{H_0,\Omega_\Lambda}\chi^2_\Sigma$ on
$\Omega_m$ is rather sharp because of the correspondent dependence
of its fraction $\chi^2_B$. This fact for $\chi^2_B$ is connected
with the contribution from the value $A(z)$ (\ref{Az}) measurements,
because  $A(z)$ is proportional to $\sqrt{\Omega_m}$ and $\chi^2_B$
is very sensitive to $\Omega_m$ values. Note that the fractions
$\chi^2_S$ and $\chi^2_H$ (in $\min\chi^2_\Sigma$) weakly depend on
$\Omega_m$.

Dependencies of $\min\chi^2_\Sigma$ on $H_0$,  $\Omega_m$ and also
$\Omega_\Lambda$, $\Omega_k$ let us calculate estimates of
acceptable values for these model parameters. They are presented
below  in Table~\ref{Estim}.

Coordinates $h=H_0/100$ and $\Omega_\Lambda$ of the minimum point
for $\chi^2_\Sigma$ depend on $\Omega_m$ in a such manner that
only for $\Omega_m\simeq0.27$ values $\Omega_\Lambda$ and
$\Omega_k$ satisfy conditions (\ref{Omk}). Note that the optimal
value  of $h$ is close to 0.7 for all $\Omega_m$ in the limits
$0<\Omega_m<1$.

\subsection{GCG model}

\begin{figure}[bh]
  \centerline{\includegraphics[scale=0.8,trim=5mm 0mm 5mm -1mm]{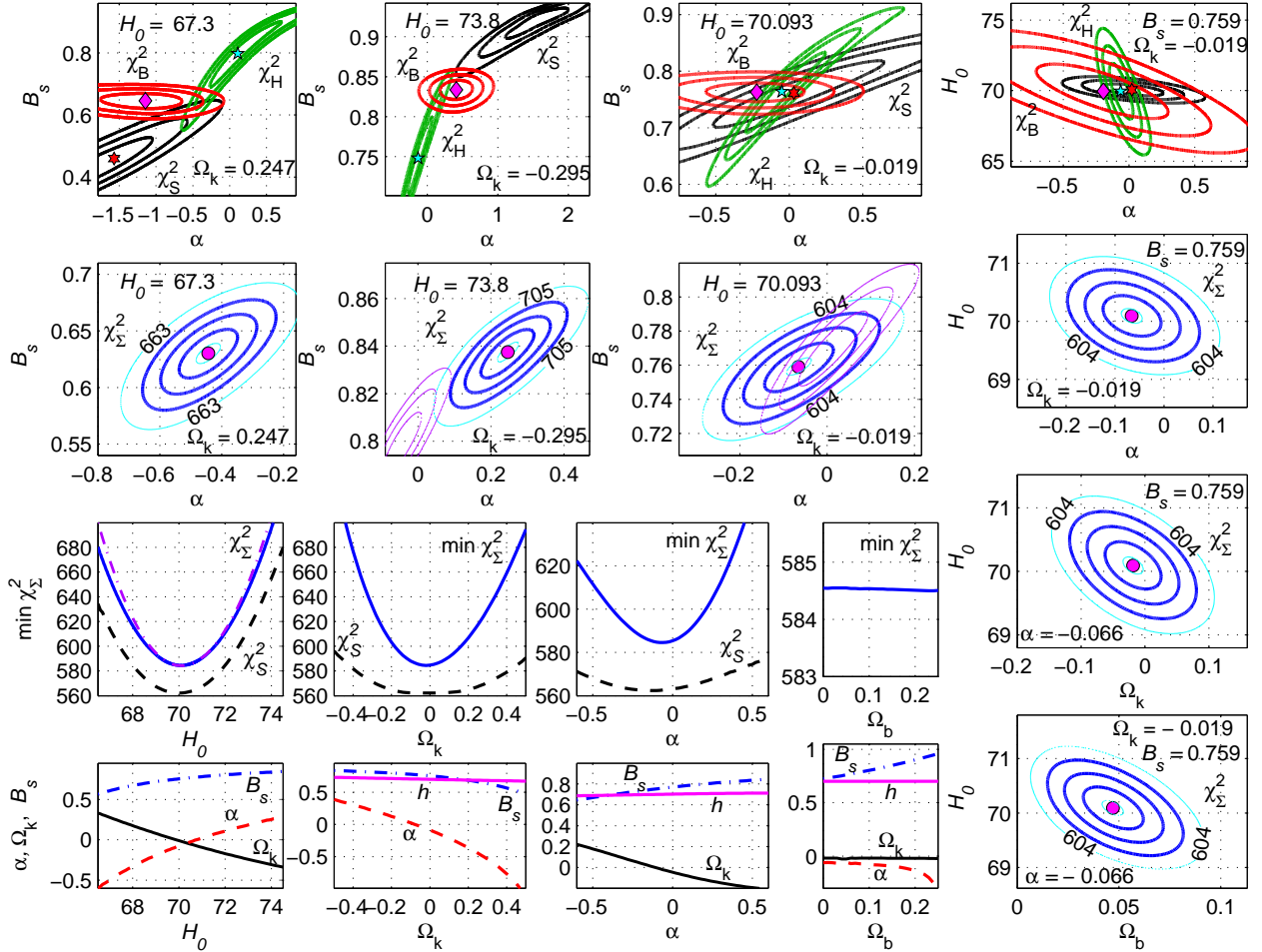}}
  \caption{\small The GCG model. For $H_0$ (\ref{H0}) and the
optimal value $H_0=70.093$ km\,c${}^{-1}$Mpc${}^{-1}$ level lines
of $\chi^2_\Sigma$ and other $\chi^2$ are presented in
$\alpha,B_s$; $\alpha,H_0$; $\Omega_k,H_0$ and $\Omega_b,H_0$
planes in notations of Fig.~\ref{F2}. In the bottom-left panels we
analyze dependence of
$\min\chi^2_\Sigma$ 
 and parameters of a minimum
point on $H_0$, $\Omega_k$, $\alpha$ and $\Omega_b$.
 }
  \label{F3}
\end{figure}

Let us apply the model with generalized Chaplygin gas  (GCG)
\cite{KamenMP01,Bento02,Makler03,LuGX10,LiangXZ11,CamposFP12,XuLu12}
to describing the same observational data for Type Ia supernovae,
$H(z)$ and BAO. We use here Eq.~(\ref{AGCG}) with the initial
condition ${\cal A}\big|_{\tau=1}=0$, so we have 5 independent
free parameters in this model: $H_0$, $\Omega_b$, $\Omega_k$,
$\alpha$ and $B_s$. However we really used only 4 free parameters,
because the fraction $\Omega_b$ may include not only baryonic but
also a part of cold dark matter. Our calculations yield that the
minimum over remaining 4 parameters $\min\limits_{H_0,\Omega_k,
\alpha,B_s}\chi^2_\Sigma$ practically does not depend on
$\Omega_b$ in the range $0\le\Omega_b\le0.25$ (see Fig.~\ref{F3}).
So in our analysis presented in Fig.~\ref{F3} (except for 3
bottom-right panels) we fixed the value
 $$ \Omega_b=0.047,
$$
 that is the simple average of the WMAP $\Omega_b=0.0464$ \cite{WMAP}
and Planck  $\Omega_b=0.0485$  \cite{Plank13} estimations.

In the GCG model $\Omega_\Lambda=0$ and $\Omega_m=1-\Omega_k$ in
accordance with Eq.~(\ref{sumOm}) and the formal definition
(\ref{Omega1}). However we should use the effective value
$\Omega_m^{eff}$ in this model, in particular, in expression
(\ref{Az}). In
Refs.~\cite{LuGX10,LiangXZ11,CamposFP12,XuLu12,LuXuWL11} the
following effective value is used
\begin{equation}
 \Omega_m^{eff}=\Omega_b+(1-\Omega_b-\Omega_k)(1-B_s)^{1/(1+\alpha)}.
 \label{Ommeff1} \end{equation}
 This value results from correspondence between the models
$\Lambda$CDM with Eq.~(\ref{ALCDM}) and GCG with  Eq.~(\ref{AGCG})
in the early universe at $z\gg1$.

But in our investigation the majority of observational data is
connected with redshifts $0<z<1$, so in Eq.~(\ref{Az}) we are to
consider the present time limit of the value
$\Omega_m^{eff}\equiv\Omega_{0m}^{eff}=\lim\limits_{z\to0}\Omega_m^{eff}$.
If we compare limits of the right hand sides of Eqs.~(\ref{ALCDM})
and (\ref{AGCG}) at $z\to0$ or ${\cal A}\to0$, we obtain another
effective value
\begin{equation}
 \Omega_m^{eff}=\Omega_b+(1-\Omega_b-\Omega_k)(1-B_s).
 \label{Ommeff2} \end{equation}

Values $\chi^2_B$ calculated with expressions (\ref{Ommeff1}) and
(\ref{Ommeff2}) are different if $\alpha\ne0$. This difference
looks like rather small if we compare minima of the sum
(\ref{chisum}) $\min\chi^2_\Sigma=\min\limits_{\Omega_k,
\alpha,B_s}\chi^2_\Sigma$ depending on $H_0$. In Fig.~\ref{F3}
this dependence with Eq.~(\ref{Ommeff2}) for $\Omega_m^{eff}$ is
the blue solid line and for the case with Eq.~(\ref{Ommeff1}) it
is the violet dash-and-dot line. We see that the lines closely
converge in the vicinity of the minimum point $H_0\simeq70$
km\,c${}^{-1}$Mpc${}^{-1}$. The  dependence
$\min\chi^2_\Sigma(H_0)$ in both cases (\ref{Ommeff1}) and
(\ref{Ommeff2}) has the sharp minimum and resembles the case of
the $\Lambda$CDM model in Fig.~\ref{F2}. The value
$\min\chi^2_\Sigma\simeq584.54$ of this minimum, its parameters in
Table~\ref{Optim}, graph of the contribution $\chi^2_S$ and
dependence on $H_0$ for parameters $\alpha,\Omega_k,B_s$ of the
minimum point in the bottom-left panel in Fig.~\ref{F3} are
presented for the case with Eq.~(\ref{Ommeff2}).

One should note that all mentioned dependencies are different for
the case (\ref{Ommeff1}), in particular, the absolute  minimum of
$\chi^2_\Sigma$ is 584.31. This difference is illustrated in the
central panels in Fig.~\ref{F3} with level lines of
$\chi^2_\Sigma(\alpha,B_s)$ for $H_0=73.8$ and 70.093
km\,c${}^{-1}$Mpc${}^{-1}$ (with the specified values $\Omega_k$,
optimal for these $H_0$). These level lines are blue for the
expression (\ref{Ommeff2}) and they are thin violet for
Eq.~(\ref{Ommeff1}). Positions of the optimal points are close
only if  $H_0$ is close to its optimal value in Table~\ref{Optim}.

We suppose that the estimation of $\chi^2_B$ with the expression
(\ref{Ommeff2}) is more adequate to the considered values $z$. So
in Table~\ref{Optim} and in other panels of Fig.~\ref{F3} we use
only Eq.~(\ref{Ommeff2}). Notations in  Fig.~\ref{F3} correspond
to Fig.~\ref{F2}.

The similar dependence of $\min\chi^2_\Sigma$ on $H_0$ for the
$\Lambda$CDM and GCG models results in unsuccessful description of
the data with $H_0=67.3$ and 73.8 km\,c${}^{-1}$Mpc${}^{-1}$ with
the corresponding optimal values $\Omega_k=0.247$ and $-0.295$.
Fig.~\ref{F3} illustrates large distances between minimum points
of $\chi^2_S$, $\chi^2_H$ and  $\chi^2_B$ in these cases. The
mentioned distances are small for the optimal values from
Table~\ref{Optim} $H_0=70.093$ km\,c${}^{-1}$Mpc${}^{-1}$ and
$\Omega_k=-0.19$. For these optimal values we present level lines
of $\chi^2_\Sigma$ in $\alpha,B_s$; $\alpha,H_0$; $\Omega_k,H_0$
and $\Omega_b,H_0$ planes. In these panels other model parameters
are fixed and specified.

When we test dependence of the minimum $\min\chi^2_\Sigma$ on
$H_0$, $\Omega_k$, $\alpha$ and $\Omega_b$ in Fig.~\ref{F3}, we
minimize this value over all other parameters (except for the
above mentioned $\Omega_b$). In particular,
$\min\chi^2_\Sigma(\Omega_k)=\min\limits_{H_0,
\alpha,B_s}\chi^2_\Sigma$, this function has the distinct minimum
near $\Omega_k\simeq0$ and resembles the dependence
$\min\chi^2_\Sigma(H_0)$. The optimal value of $H_0$ or
$h=H_0/100$ is practically constant and close to $h\simeq0.7$ if
we vary $\Omega_k$, $\alpha$ or $\Omega_b$.
 As
mentioned above the dependence of $\min\chi^2_\Sigma$ on
$\Omega_b$ is very weak, so we fixed in our previous analysis
$\Omega_b=0.047$.

For the graph
$\min\chi^2_\Sigma(\alpha)=\min\limits_{H_0,\Omega_k,B_s}\chi^2_\Sigma$
the correspondent minimum is achieved if $\alpha$
 is negative: $\alpha=-0.066$ (see Table~\ref{Optim}). In the GCG model
this parameter is connected with the square of adiabatic sound
speed \cite{Makler03,CamposFP12,XuLu12}
\begin{equation}
 c_s^2 = \frac{\delta p}{\delta\rho}=-\alpha \frac{p}\rho.
 \label{cs2} \end{equation}

If we accept the restriction $\alpha\ge0$ (equivalent to
$c_s^2\ge0$) in our investigation with the mentioned observational
data, we obtain the optimal value $\alpha=0$ and the GCG model
will be reduced to the $\Lambda$CDM model with
$\Omega_\Lambda=B=B_s(1-\Omega_b-\Omega_k)$. The dependence of
$\min\chi^2_\Sigma$ and other parameters on $\alpha$ in
Fig.~\ref{F3} show that for  $\alpha=0$ we have
$\min\chi^2_\Sigma\simeq585.35$ and the optimal values of $H_0$,
$\Omega_k$, $\Omega_\Lambda=B$ corresponding to the $\Lambda$CDM
model in Table~\ref{Optim}.

\begin{table}[hb]
\caption{Optimal values of model parameters ($\Omega_b=0.047$, for
the GCG model $\Omega_m=\Omega_m^{eff}$ (\ref{Ommeff2})).}
\begin{center}
\begin{tabular}{||l||c||c|c|l||}  \hline
 Model       &$\min\chi^2_\Sigma$&$H_0$ &$\Omega_m$& other parameters  \\ \hline
 $\Lambda$CDM& 585.35& 70.262& 0.276&$\Omega_\Lambda=0.769,\;\; \Omega_k=-0.045$ \\ \hline
 GCG         & 584.54& 70.093& 0.277&$\Omega_k=-0.019,\;\;\alpha=-0.066,\;\;B_s=0.759$ \\ \hline
 PCS, $d=1$  & 588.41& 69.52 & 0.286 &$\Omega_k=-0.040,\;\;\alpha=-0.256,\;\;B=2.067$\\ \hline
 PCS, $d=2$  & 591.10& 69.49 & 0.288 &$\Omega_k=-0.017,\;\;\alpha=-0.372,\;\;B=1.599$\\ \hline
 PCS, $d=3$  & 592.18& 69.34 & 0.288 &$\Omega_k=-0.027,\;\;\alpha=-0.431,\;\;B=1.461$\\ \hline
 PCS, $d=6$  & 592.56& 69.29 & 0.289 &$\Omega_k=-0.029,\;\;\alpha=-0.493,\;\;B=1.302$\\ \hline
 \end{tabular}
\end{center}
 \label{Optim}\end{table}

\subsection{PCS model}

The multidimensional gravitational model of I. Pahwa, D.~Choudhury
and T.R.~Seshadri \cite{PahwaChS} has the set of model parameters
$H_0$, $\Omega_b$, $\Omega_m$, $\Omega_k$, $\alpha$, $B$ similar
to the GCG model, but also it has the additional integer-valued
parameter $d$ (the number of extra dimensions). Our analysis
demonstrated that the value $d=1$ is the most preferable for
describing the observational data for supernovae, BAO and $H(z)$.

\begin{figure}[ht]
  \centerline{\includegraphics[scale=0.8,trim=5mm 0mm 5mm -1mm]{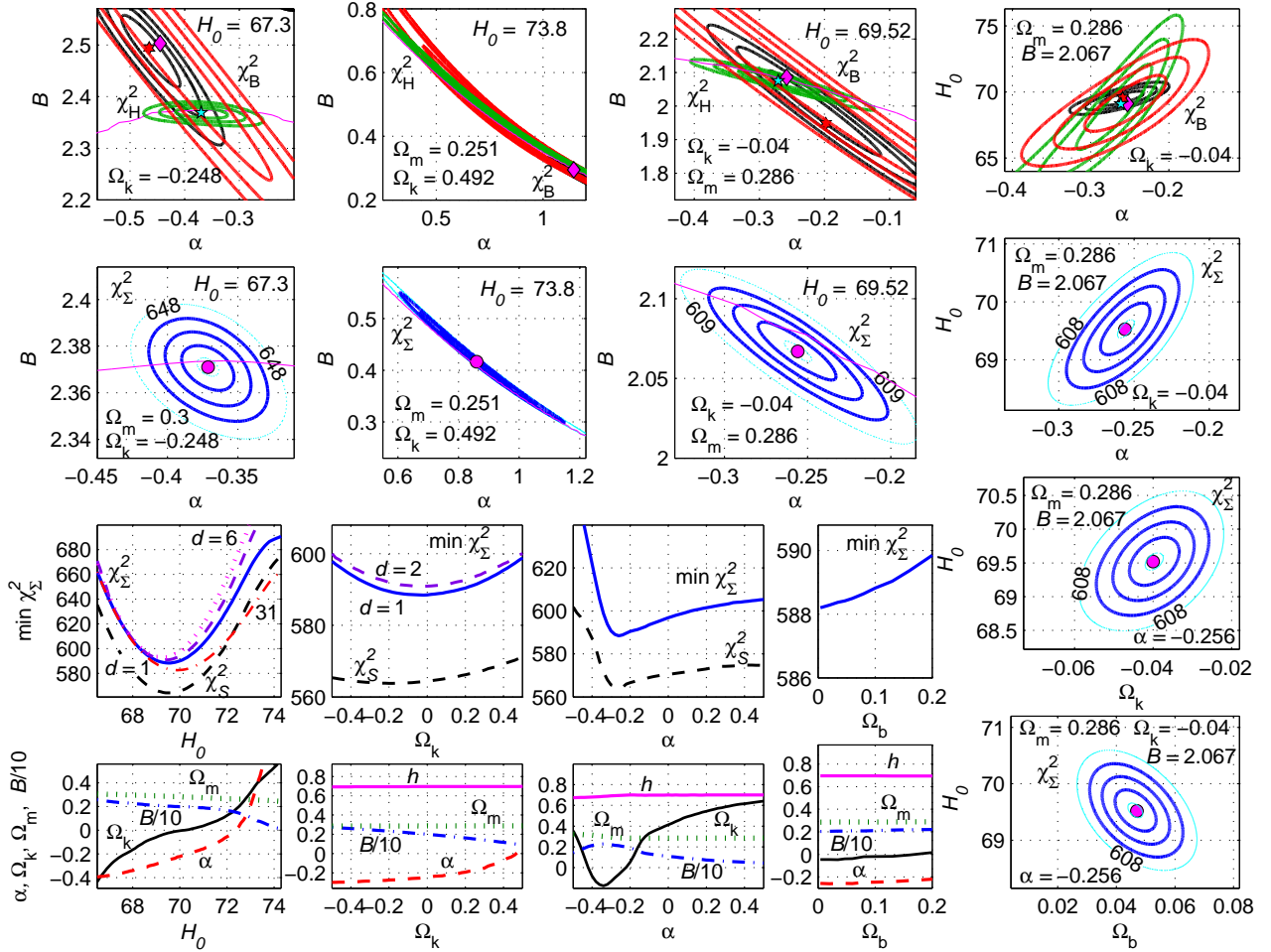}}
  \caption{\small The PCS model with $d=1$. Notations and panels
correspond to Fig.~\ref{F3}, in particular, in the bottom-left
 panels we analyze dependence of
$\min\chi^2_\Sigma$ 
 and parameters of a minimum
point on $H_0$, $\Omega_k$, $\alpha$ and $\Omega_b$.
 }
  \label{F4}
\end{figure}

So it is the case $d=1$ that we present in almost all panels of
Fig.~\ref{F4} (except for 2 panels with dependencies of
$\min\chi^2_\Sigma$ on $H_0$ and $\Omega_k$). We use the
similarity of model parameters for the GCG and PCS models draw in
Fig.~\ref{F4} the same graphs and level lines for the PCS model as
in Fig.~\ref{F3} in correspondent panels. Colors of correspondent
lines also coincide. Naturally we use in Fig.~\ref{F4} the value
$B$ instead of $B_s$.

The minimum $\min\chi^2_\Sigma$ (over
all other parameters) increases when the baryon fraction $\Omega_b$ grows.
 This dependence is more distinct than in the GCG case
 (Fig.~\ref{F3}), but it is also rather weak for small $\Omega_b$.
So for the multidimensional model PCS we also fix $\Omega_b=0.047$
and really use only 5 remaining parameters $H_0$, $\Omega_m$,
$\Omega_k$, $\alpha$, $B$. The value $\Omega_b=0.047$ is fixed in
all panels of Fig.~\ref{F4} like for  Fig.~\ref{F3} (except for 3
bottom-right panels).

The dependence of
$\min\chi^2_\Sigma=\min\limits_{\Omega_m,\Omega_k,
\alpha,B}\chi^2_\Sigma$ on $H_0$ has the distinct minimum at
$H_0\simeq69.52$ for $d=1$ (the solid blue line here and in panels
of this row). The similar behavior takes place for $d=2$ (the
violet dashed line) and for $d=6$ (the purple dots). The minimal
value $\min\chi^2_\Sigma\simeq588.41$ for $d=1$ is larger than for
the $\Lambda$CDM and GCG models and for $d\ge2$ the minima are
still worse (see Table~\ref{Optim}).

These bad results for the PCS model are connected with description
of the $H(z)$ recent data with high $z$ ($z>2$ in
Table~\ref{AT1}). When we excluded 3 data points
\cite{Busca12,Delubac14,Font-Ribera13} for $H(z)$ with $z\ge2.3$,
we obtained absolutely other results presented below in
Table~\ref{Optim31}.

In Fig.~\ref{F4} all level lines and graphs correspond to the
whole $H(z)$ data with $N_H=34$ points. But only one except is
done for the dependence of $\min\chi^2_\Sigma$ on $H_0$ for $d=1$:
here $N_H=31$, this graph is shown as the red dash-and-dot line.
The minimum value for this line $\min\chi^2_\Sigma\simeq582.68$ is
in Table~\ref{Optim31}.

Level lines of functions $\chi^2$ are shown in Fig.~\ref{F4} in
the same panels as for the GCG model in Fig.~\ref{F3}, in
 particular, for the values (\ref{H0})
 $H_0=67.3$, $73.8$  and the optimal
 value $69.52$ km\,c${}^{-1}$Mpc${}^{-1}$. If $H_0$ is too large,
the domain of acceptable level of $\chi^2_\Sigma$ becomes very
narrow. One should note that for all level lines we change only
two parameters, all remaining model parameters are fixed (they are
from Table~\ref{Optim} or optimal for a given $H_0$).

In 6 top-left panels with the $\alpha,B$ plane we draw  thin
purple lines bounding the domain of regular solutions (below these
lines). The upper domain (for larger $B$) consists of singular
solutions, they have singularities in the past with infinite value
of density $\rho$ corresponding to nonzero value of the scale
factor $a$ \cite{GrSh13}. These solutions are nonphysical and
should be excluded. It is interesting that the optimal solutions
in Fig.~\ref{F4} and in Tables~\ref{Optim} and \ref{Optim31} are
near this border, but they are regular and describe the standard
Big Bang $\rho\to\infty\;\Leftrightarrow\;a\to0$ with dynamical
compactification of extra dimensions.

\section{Conclusion}

We considered how the $\Lambda$CDM, GCG and PCS models describe
the observational data for type Ia supernovae, BAO  and $H(z)$
\cite{SNTable}, Tables~\ref{AT2}, \ref{AT1}. These observations
distinctly restrict acceptable values for the Hubble constant
$H_0$ and other parameters of the mentioned models. We used our
calculations for dependance $\min\chi^2_\Sigma(p)$, where the
absolute minimum (over other parameters) of the value
(\ref{chisum}) $\chi^2_\Sigma$ depend on a fixed parameter $p$. On
the base of these calculations (presented partially in
Figs.~\ref{F2}, \ref{F3}, \ref{F4}) we obtained the following
$1\sigma$ estimates for parameters of the $\Lambda$CDM, GCG and
PCS ($d=1$) models:
  \begin{table}[h]
\caption{$1\sigma$ estimates of model parameters ($\Omega_b=0.047$
in the GCG and PCS models).} 
\begin{tabular}{||l||c||c|c|l||}  \hline
 Model       &$\min\chi^2_\Sigma$&$H_0$&$\Omega_k$&other parameters  \\ \hline
 $\Lambda$CDM$\!$& 585.35& $70.262\pm0.319$&$-0.04\pm0.032$& $\Omega_m$=$\,0.276_{-0.008}^{+0.009},
\rule[-0.5em]{0mm}{1.6em}\; \;\Omega_\Lambda$=$\,0.769\pm0.029$ \\
\hline
 GCG         & 584.54& $70.093\pm0.369$&$-0.019\pm0.045$ &$\alpha\!=
 \!-0.066_{-0.074}^{+0.072},\;\; B_s$=\,$0.759_{-0.016}^{+0.015}$
 \rule[-0.5em]{0mm}{1.6em} \\ \hline
PCS, $d=1$\rule{0mm}{1.2em}& 588.41& $69.523_{-0.350}^{+0.366}$ &
$-0.04\pm0.045$&$\Omega_m$=$\,0.286\pm0.010,\,\;\alpha=-0.256_{-0.03}^{+0.032}$\\
\hline
\end{tabular}
 \label{Estim}\end{table}

Our estimates for the $\Lambda$CDM model are in agreement with the
WMAP  observational restrictions (\ref{Omk}) on $\Omega_m$,
$\Omega_\Lambda$, $\Omega_k$ \cite{WMAP}, but they are in tension
with the Planck data \cite{Plank13}. This fact is connected with
 too low value $H_0=67.3$
km\,c${}^{-1}$Mpc${}^{-1}$ (\ref{H0}) in the Planck survey
\cite{Plank13}.

For the GCG model $\min\chi^2_\Sigma$ is slightly better and our
limitations on $H_0$ and $\Omega_k$ in Table~\ref{Estim} are
rather close to the $\Lambda$CDM case. However, if we require
$\alpha\ge0$ in accordance with Eq.~(\ref{cs2}) and
Refs.~\cite{CamposFP12,XuLu12}, the GCG model with the optimal
value $\alpha=0$  will be reduced to the $\Lambda$CDM model with
its optimal parameters in Tables~\ref{Optim}, \ref{Estim} and the
same $\min\chi^2_\Sigma$.

Values $\chi^2_B$ and $\chi^2_\Sigma$ for the GCG model
essentially depend on the expression for $\Omega_m^{eff}$
(\ref{Ommeff1}) or (\ref{Ommeff2}). But the optimal parameters in
Table~\ref{Optim} for these expressions are rather close.

We mentioned above that the multidimensional  model PCS is less
effective in description of the considered observational data, and
that the main problem of this model is connected with the $H(z)$
recent data with high $z$ ($z>2$). We excluded 3 $H(z)$ data
points \cite{Busca12,Delubac14,Font-Ribera13} with $z=2.3$, 2.34,
 2.36 and for remaining $N_H=31$ points of $H(z)$ and the same
SN and BAO data from \cite{SNTable}, Table~\ref{AT2}. we
calculated $\min\chi^2_\Sigma$ and optimal values of model
parameters presented here in Table~\ref{Optim31}.

\begin{table}[hb]
\caption{Optimal values of model parameters  for $\Omega_b=0.047$
and $N_H=31$ $H(z)$ data points with $z<2$.}
\begin{center}
\begin{tabular}{||l||c||c|c|l||}  \hline
 Model       &$\min\chi^2_\Sigma$&$H_0$ &$\Omega_m$& other parameters  \\ \hline
 $\Lambda$CDM& 583.71& 70.12 & 0.281&$\Omega_\Lambda=0.751,\;\; \Omega_k=-0.032$ \\ \hline
 GCG         & 583.70& 70.11 & 0.291&$\Omega_k=-0.046,\;\;\alpha=-0.028,\;\;B_s=0.756$ \\ \hline
 PCS, $d=1$  & 582.68& 69.89 & 0.281 &$\Omega_k=-0.114,\;\;\alpha=-0.174,\;\;B=2.078$\\ \hline
 PCS, $d=2$  & 582.93& 69.82 & 0.282 &$\Omega_k=-0.118,\;\;\alpha=-0.290,\;\;B=1.616$\\ \hline
 PCS, $d=6$  & 583.23& 69.78 & 0.282 &$\Omega_k=-0.126,\;\;\alpha=-0.398,\;\;B=1.291$\\ \hline
 \end{tabular}
\end{center}
 \label{Optim31}\end{table}

We see that the model PCS \cite{PahwaChS} describes the reduced
set of data with $z<2$ better than other models. The best fit is
for $d=1$, the optimal value of $H_0$ close to $70$
km\,c${}^{-1}$Mpc${}^{-1}$.

This example demonstrates that predictions of any cosmological
model essentially depend on data selection. Moreover, there is the
important problem of model dependence (in addition to mutual
dependence) of observational data, in particular, data in
Tables~\ref{AT2}, \ref{AT1}.

Leaving the last problem beyond this paper, we can conclude that
the considered observations of type Ia supernovae \cite{SNTable},
BAO  (Table~\ref{AT2}) and the Hubble parameter $H(z)$
(Table~\ref{AT1}) confirm effectiveness of the $\Lambda$CDM model,
but they do not deny other models. The important argument in favor
of the $\Lambda$CDM model is its small number $N_p$ of model
parameters (degrees of freedom). This number is part of
information criteria of model selection statistics, in particular,
the Akaike information criterion is \cite{ShiHL12} $AIC =
\min\chi^2_\Sigma + 2N_p$. This criterion supports the leading
position of the $\Lambda$CDM model.

\appendix
\section{Appendix}\label{App}

\begin{table}[h]
\caption{Values of $d_z(z)=r_s(z_d)/D_V(z)$ (\ref{dz}) and $A(z)$
(\ref{Az}) with corresponding errors
\cite{WMAP,BlakeBAO11,Chuang13}}
\begin{center}
\begin{tabular}{||l|l|l|l|l|c||}  \hline
 $z$  & $d_z(z)$ &$\sigma_d$    & ${ A}(z)$ & $\sigma_A$  & Refs\\ \hline
0.106 &0.336  & 0.015 & 0.526& 0.028&\cite{WMAP} \\ \hline
0.20  &0.1905 & 0.0061& 0.488& 0.016& \cite{WMAP} \\ \hline
0.35  &0.1097 & 0.0036& 0.484& 0.016& \cite{WMAP}   \\ \hline
0.44  &0.0916 & 0.0071& 0.474& 0.034& \cite{BlakeBAO11} \\ \hline
0.57 &0.07315 & 0.0012& 0.436& 0.017& \cite{WMAP,Chuang13}   \\ \hline
0.60 &0.0726  & 0.0034& 0.442& 0.020& \cite{BlakeBAO11} \\ \hline
0.73 &0.0592  & 0.0032& 0.424& 0.021& \cite{BlakeBAO11} \\ \hline
 \end{tabular}
\end{center}
 \label{AT2}\end{table}

Measurements of $d_z(z)$ and ${ A}(z)$ in Ref.~\cite{BlakeBAO11}
are not independent, they are described with the following
elements of covariance matrices $C_d^{-1}=||c^d_{ij}||$ and
$C_A^{-1}=||c^A_{ij}||$ in Eq.~(\ref{chiB})
\cite{WMAP,BlakeBAO11}:
$$\begin{array}{lll}
c^d_{44}=24532.1,& c^d_{46}=-25137.7,&  c^d_{47}=12099.1,\\
c^d_{66}=134598.4,& c^d_{67}=-64783.9,& c^d_{77}=128837.6;\\
 c^A_{44}=1040.3,& c^A_{46}=-807.5,& c^A_{47}=336.8,\\
 c^A_{66}=3720.3,& c^A_{67}=-1551.9,& c^A_{77}=2914.9.
 \end{array}$$
 These matrices are symmetric ones, their remaining  elements are
 $c_{ii}=1/\sigma_i^2$, $c_{ij}=0$, $i\ne j$.

\bigskip

\begin{table}[h]
\caption{Values of the Hubble parameter $H(z)$ with errors
$\sigma_H$ from
Refs.~\cite{Simon05,Stern10,Moresco12,Blake12,Zhang12,Busca12,Chuang12,Gazta09,Anderson13,Oka13,Delubac14,Font-Ribera13}}
\begin{center}
\begin{tabular}{||l|l|l|c||l|l|l|c||}  \hline
 $z$  & $H(z)$ &$\sigma_H$  & Refs &   $z$ & $H(z)$ & $\sigma_H$  & Refs\\ \hline
0.070 & 69  & 19.6& \cite{Zhang12} & 0.57  &
92.9&7.855&\cite{Anderson13} \\ \hline 0.090 & 69  & 12  &
\cite{Simon05} & 0.593 & 104 & 13  & \cite{Moresco12} \\ \hline
0.120 & 68.6& 26.2& \cite{Zhang12} & 0.600 & 87.9& 6.1 &
\cite{Blake12}   \\ \hline 0.170 & 83  & 8   & \cite{Simon05} &
0.680 & 92  & 8;  & \cite{Moresco12} \\ \hline 0.179 & 75  & 4
&\cite{Moresco12}& 0.730 & 97.3& 7.0 & \cite{Blake12}   \\ \hline
0.199 & 75  & 5   &\cite{Moresco12}& 0.781 & 105 & 12  &
\cite{Moresco12} \\ \hline 0.200 & 72.9& 29.6& \cite{Zhang12} &
0.875 & 125 & 17  & \cite{Moresco12} \\ \hline 0.240 &79.69& 2.65&
\cite{Gazta09} & 0.880 & 90  & 40  & \cite{Stern10}   \\ \hline
0.270 & 77  & 14  & \cite{Simon05} & 0.900 & 117 & 23  &
\cite{Simon05}   \\ \hline 0.280 & 88.8& 36.6& \cite{Zhang12} &
1.037 & 154 & 20  & \cite{Moresco12} \\ \hline 0.300 & 81.7& 6.22&
\cite{Oka13}   & 1.300 & 168 & 17  & \cite{Simon05}   \\ \hline
0.350 & 82.7& 8.4 & \cite{Chuang12}& 1.430 & 177 & 18  &
\cite{Simon05}   \\ \hline 0.352 & 83  & 14  &\cite{Moresco12}&
1.530 & 140 & 14  & \cite{Simon05}   \\ \hline 0.400 & 95  & 17  &
\cite{Simon05} & 1.750 & 202 & 40  & \cite{Simon05}   \\ \hline
0.430 &86.45& 3.68& \cite{Gazta09} & 2.300 & 224 & 8   &
\cite{Busca12}   \\ \hline 0.440 & 82.6& 7.8 & \cite{Blake12} &
2.340 & 222 & 7   & \cite{Delubac14} \\ \hline 0.480 & 97  & 62  &
\cite{Stern10} & 2.360 & 226 & 8   & \cite{Font-Ribera13}\\ \hline
 \end{tabular}
\end{center}
 \label{AT1}\end{table}
\medskip     

\acknowledgments
 G.S. would like to acknowledge the support of
the Ministry of education and science of Russia (grant No.
1.476.2011).


\begin{thebibliography}{22}

\bibitem{Riess98}
 A.G. Riess et al., \emph{Observational
Evidence from Supernovae for an Accelerating Universe and a Cosmological
Constant}, \emph{Astron. J.} {\bf 116} (1998) 1009 
[astro-ph/9805201].

\bibitem{Perl99} S. Perlmutter et al., \emph{Measurements of Omega and Lambda
from 42 high redshift supernovae}, \emph{Astrophys. J.} {\bf 517} (1999) 565 
[astro-ph/9812133].

\bibitem{SNTable}
N. Suzuki et al., \emph{The Hubble Space Telescope Cluster Supernova Survey: V.
Improving the Dark Energy Constraints Above z>1 and Building an
Early-Type-Hosted Supernova Sample}, \emph{Astrophys. J.} {\bf 746} (2012) 85
[arXiv:1105.3470]. 

\bibitem{WeinbergAcc12}
 D.H. Weinberg et al., \emph{Observational Probes of Cosmic Acceleration}, \emph{ Physics Reports} {\bf 530} (2013)  87  [arXiv:1201.2434].

\bibitem{WMAP} G. Hinshaw et al., \emph{Nine-year Wilkinson Microwave Anisotropy Probe (WMAP) Observations:  Cosmological Parameters Results},
\emph{Astrophysical Journal Suppl.} {\bf 208} (2013)  19 [arXiv:1212.5226].

\bibitem{Eisen05}
 D.J. Eisenstein et al., \emph{Detection of the baryon acoustic peak in the large-scale correlation function of
SDSS luminous red galaxies}, \emph{Astrophys. J.} {\bf 633(2)} (2005) 560 
[astro-ph/0501171].

\bibitem{SDSS} B.A. Reid  et al., \emph{Cosmological constraints from the clustering of the Sloan Digital
 Sky Survey DR7 luminous red galaxies}, \emph{Mon. Not. Roy. Astron. Soc.} {\bf 404} (2010) 60
[arXiv:0907.1659]. 


\bibitem{Plank13}  P.A.R. Ade et al., \emph{Planck 2013 results. XVI. Cosmological
parameters}, arXiv:1303.5076.  

\bibitem{Simon05}
 J. Simon, L. Verde and R. Jimenez, \emph{Constraints on the redshift dependence of the dark energy
potential}, \emph{Phys. Rev. {\bf D}} {\bf 71} (2005) 123001
[astro-ph/0412269].


\bibitem{Stern10}
D. Stern,  R. Jimenez,  L. Verde, M. Kamionkowski and S. A. Stanford,
\emph{Cosmic chronometers: constraining the equation of state of dark energy.
I: $H(z)$ measurements}, \emph{J. of Cosmology and Astropart. Phys.} {\bf 02}
(2010) 008 [arXiv:0907.3149].

\bibitem{Moresco12}
M. Moresco et al., \emph{Improved constraints on the expansion rate of the
Universe up to $z$~1.1 from the spectroscopic evolution of cosmic
chronometers}, \emph{J. of Cosmology and Astropart. Phys.} {\bf 8} (2012) 006
[arXiv:1201.3609].


\bibitem{Blake12}
C. Blake et al., \emph{The WiggleZ Dark Energy Survey: Joint measurements of
the expansion and growth history at
$z < 1$ }, \emph{Mon. Not. Roy. Astron. Soc.} {\bf 425(1)} (2012) 405 
[arXiv:1204.3674]. 

\bibitem{Zhang12}
 C. Zhang et al., \emph{Four New Observational H(z)  Data From Luminous
Red Galaxies Sloan Digital Sky Survey Data Release Seven}, arXiv:1207.4541. 

\bibitem{Busca12}
 N.G. Busca et al., \emph{Baryon Acoustic Oscillations in the Ly$\alpha$ forest of BOSS quasars}, \emph{Astron. and Astrop.}
 {\bf 552} (2013) A96 [arXiv:1211.2616].

\bibitem{Chuang12}
C.H. Chuang and Y. Wang, \emph{Modeling the Anisotropic Two-Point
Galaxy Correlation Function on Small Scales and Improved
Measurements of $H(z)$, $D_A(z)$, and $f(z)\sigma_8(z)$ from the
Sloan Digital Sky Survey DR7 Luminous Red Galaxies}, \emph{Mon. Not.
Roy. Astron. Soc.} {\bf 435(1)}
(2013) 255 
[arXiv:1209.0210].

\bibitem{Gazta09}
E. Gazta\~naga,  A. Cabre, L. Hui, \emph{Clustering of Luminous Red Galaxies
IV: Baryon Acoustic Peak in the Line-of-Sight Direction
and a Direct Measurement of $H(z)$}, \emph{Mon. Not. Roy. Astron. Soc.} {\bf 399(3)} (2009) 1663 
[arXiv:0807.3551].

\bibitem{Anderson13}
L. Anderson et al., \emph{The clustering of galaxies in the SDSS-III Baryon
Oscillation Spectroscopic Survey: Measuring $D_A$ and $H$ at $z = 0.57$ from
the Baryon Acoustic Peak in the Data Release 9 Spectroscopic Galaxy Sample},
 \emph{Mon. Not. Roy. Astron. Soc.} {\bf 439(1)} (2014)
 83 
[arXiv:1303.4666].

\bibitem{Oka13}
A. Oka et al., \emph{Simultaneous constraints on the growth of
structure and cosmic expansion from the multipole power spectra of
the SDSS DR7 LRG sample},  \emph{Mon. Not. Roy. Astron. Soc.} {\bf
439(3)} (2014) 2515 
[arXiv:1310.2820].

\bibitem{Delubac14}
T. Delubac et al., \emph{Baryon Acoustic Oscillations in the
Ly$\alpha$ forest of BOSS DR11 quasars}, arXiv:1404.1801.

\bibitem{Font-Ribera13}
A. Font-Ribera et al., \emph{Quasar-Lyman $\alpha$ Forest
Cross-Correlation from BOSS DR11: Baryon Acoustic Oscillations},
\emph{J. of Cosmology and Astroparticle Phys.} {\bf 05} (2014) 027
[arXiv:1311.1767].

\bibitem{BlakeBAO11}
C. Blake et al., \emph{The WiggleZ dark energy survey: mapping the
distance redshift relation with baryon acoustic oscillations}, \emph{Mon. Not. Roy. Astron. Soc.} {\bf 418(3)} (2011) 1707 
[arXiv:1108.2635]. 

\bibitem{Chuang13}
C-H. Chuang et al., \emph{The clustering of galaxies in the SDSS-III
Baryon Oscillation Spectroscopic Survey: single-probe measurements
and the strong power of $f(z)\sigma_8(z)$ on constraining dark
energy},  \emph{Mon. Not. Roy. Astron. Soc.} {\bf 433(4)}
(2013) 3559 
[arXiv:1303.4486]. 

\bibitem{Clifton}
T. Clifton, P.G. Ferreira, A. Padilla and C. Skordis, \emph{Modified Gravity
and Cosmology},
\emph{Physics Reports} {\bf 513} (2012) 1 [arXiv:1106.2476]. 

\bibitem{Bamba12}
K. Bamba, S. Capozziello, S. Nojiri and S.D.~Odintsov, \emph{Dark energy
cosmology: the equivalent description via different theoretical models and
cosmography tests},
\emph{Astrophys. and Space Science} {\bf 342} (2012) 155 
 [arXiv:1205.3421].

\bibitem{CopelandST06}
E.J. Copeland, M. Sami and S. Tsujikawa, \emph{Dynamics of dark energy},
\emph{Int. J. Mod. Phys. {\bf D}} {\bf 15} (2006) 1753 [hep-th/0603057].

\bibitem{Kunz12}
M. Kunz, \emph{The phenomenological approach to modeling the dark energy},
\emph{Comptes rendus - Physique} {\bf 13} (2012) 539,  
[arXiv:1204.5482]. 

\bibitem{SotiriouF}
T.P. Sotiriou and V. Faraoni, \emph{f(R) theories of gravity}, \emph{Rev. of
Modern Phys.} {\bf 82} (2010) 451 [arXiv:0805.1726].

\bibitem{NojOdinFR}
S. Nojiri and S. D. Odintsov, \emph{Unified cosmic history in modified gravity:
from F(R) theory to Lorentz non-invariant models}, \emph{Phys. Rept.} {\bf 505} (2011) 59 
[arxiv:1011.0544].

\bibitem{CaldwellDS98}
R. R. Caldwell, R. Dave and P. J. Steinhardt, \emph{Cosmological imprint of an
energy component with general equation of state}, \emph{Phys. Rev. Lett.} {\bf
80} (1998) 1582 [astro-ph/9708069].

\bibitem{KhouryA04}
J. Khoury and A. Weltman, \emph{Chameleon cosmology}, \emph{Phys. Rev. {\bf D}}
{\bf 69} (2004) 044026 [astro-ph/0309411].

\bibitem{KamenMP01}
A.Y. Kamenshchik, U. Moschella and V. Pasquier \emph{An alternative to
quintessence},
\emph{Phys. Lett. {\bf B}} {\bf 511(2-4)} (2001) 265 
[arXiv:gr-qc/0103004].

\bibitem{Bento02}
M.C. Bento, O. Bertolami and  A.A. Sen, \emph{Generalized Chaplygin gas,
accelerated expansion, and dark-energy-matter unification}, {Phys. Rev. {\bf
D}} {\bf 66(4)} (2002) 043507 [arXiv:gr-qc/0202064].

\bibitem{Makler03}
M. Makler, S.Q. de Oliveira and I. Waga, \emph{Constraints on the generalized
Chaplygin gas from supernovae observations}, \emph{Phys.
Lett. {\bf B}} {\bf 555(1-2)} (2003) 1 
[arXiv:astro-ph/0209486].


\bibitem{LuGX10}
J. Lu, Y. Gui and L. Xu, \emph{Observational constraint on
generalized
Chaplygin gas model}, \emph{Eur. Phys. J. C.} {\bf 63} (2009) 349 
[arXiv:1004.3365]. 

\bibitem{LiangXZ11}
N. Liang, L. Xu and Z. Zhu, \emph{Constraints on the generalized
Chaplygin gas model including gamma-ray bursts via a Markov Chain
Monte Carlo approach}, \emph{Astron. Astrophys.} {\bf 527} (2011)
A11 [arXiv:1009.6059].

\bibitem{CamposFP12}
J.P. Campos, J.C. Fabris, R. Perez, O.F. Piattella and H. Velten, \emph{Does
Chaplygin gas have salvation?} \emph{Eur. Phys. J. C.} {\bf 73}
(2013) 2357 [arXiv:1212.4136].

\bibitem{XuLu12}
L. Xu, J. Lu and Y. Wang, \emph{Revisiting Generalized Chaplygin Gas as a
Unified Dark Matter and Dark Energy Model}, \emph{Eur. Phys. J. C.} {\bf 72}
(2012) 1883 [arXiv:1204.4798].


\bibitem{LuXuWL11}
J. Lu, L. Xu, Y. Wu, and M. Liu, \emph{Combined constraints on
modified Chaplygin gas model from cosmological observed
data: Markov Chain Monte Carlo approach},  \emph{Gen. Rel. Grav.}
 {\bf 43} (2011) 819 [arXiv:1105.1870].

\bibitem{PaulTh13}
B.C. Paul and P. Thakur, \emph{Observational constraints on modified
Chaplygin gas from cosmic growth}, \emph{J. of Cosmology and Astropart. Phys.} {\bf 11} (2013) 052 [arXiv:1306.4808]. 

\bibitem{Mohammedi02}
N. Mohammedi, \emph{Dynamical Compactification, Standard Cosmology and the
Accelerating Universe}, \emph{Phys. Rev.{\bf D}} {\bf 65} (2002) 104018
[hep-th/0202119].

\bibitem{Darabi03}
 F. Darabi, \emph{Accelerating Universe and Dynamical Compactification of Extra Dimensions},
\emph{Class. Quant. Grav.} {\bf 20} (2003) 3385 [gr-qc/0301075].

\bibitem{BringmannEG03}
T. Bringmann, M. Eriksson and M. Gustafsson, \emph{Cosmological Evolution of
Homogeneous Universal Extra Dimensions}, \emph{Phys. Rev. {\bf D}} {\bf 68}
(2003) 063516 [astro-ph/0303497].

\bibitem{PanigrahiZhCh06}
D. Panigrahi, Y.Z. Zhang and S. Chatterjee, \emph{Accelerating Universe as
Window for Extra Dimensions}, \emph{Int. J. Mod. Phys. {\bf A}} {\bf21} (2006)
6491 [gr-qc/0604079].

\bibitem{MiddleSt11}
C. A. Middleton  and E. Stanley, \emph{Anisotropic evolution of 5D
Friedmann-Robertson-Walker spacetime},
\emph{Phys. Rev. {\bf D}} {\bf 84} (2011) 085013 [arXiv:1107.1828]. 

\bibitem{FarajollahiA10}
H. Farajollahi and H. Amiri, \emph{5D noncompact Kaluza -Klein cosmology in the
presence of Null perfect fluid},
\emph{Int. J. Mod. Phys. {\bf D}} {\bf19} (2010) 1823 [arXiv:1005.3140]. 

\bibitem{PahwaChS}
 I. Pahwa, D. Choudhury and T. R. Seshadri, \emph{Late-time acceleration in Higher Dimensional Cosmology},
\emph{ J. of Cosmology and Astroparticle Phys.} {\bf 09} (2011) 015 [arXiv:1104.1925]. 


\bibitem{GrSh13}
 O.A. Grigorieva and G.S. Sharov, \emph{Multidimensional gravitational
model with anisotropic pressure}, \emph{Intern. Journal of Modern Physics {\bf D}} {\bf 22} (2013) 1350075 [arXiv:1211.4992]. 

\bibitem{Riess11}
A.G. Riess et al., \emph{A 3\% Solution: Determination of the Hubble Constant
with the Hubble Space Telescope and Wide Field Camera 3}, \emph{Astrophys. J.}
{\bf 730(2)} (2011) 119 [arXiv:1103.2976].


\bibitem{Tonry03}  et al. 2003, ApJ, 594, 1
 J.L. Tonry  et al.,  \emph{Cosmological Results from High-z Supernovae},
 \emph{Astrophys. J.}  {\bf 594}
(2003) 1, arXiv:astro-ph/0305008

\bibitem{Knop03}.
 R.A. Knop et al., \emph{New Constraints on
$\Omega_m$, $\Omega_\Lambda$, and $w$ from an Independent Set of
Eleven High-Redshift Supernovae Observed with HST1},
\emph{Astrophys. J.}  {\bf 598} (2003) 102,
arXiv:astro-ph/0309368.


\bibitem{Kowalski08}
 M. Kowalski  et al.,  \emph{Improved cosmological constraints from new, old and
combined supernova datasets},  \emph{Astrophys. J.}  {\bf 686}
(2008) 749 [arXiv:0804.4142].

\bibitem{ShiHL12}
 K. Shi, Y.F. Huang and T. Lu,  \emph{A comprehensive comparison of
cosmological models from the latest observational data},
\emph{Monthly Notices Roy. Astronom. Soc.}  {\bf 426} (2012) 2452
[arXiv:1207.5875]. 



\bibitem{FarooqMR13}
O. Farooq, D. Mania and B. Ratra, \emph{Hubble parameter measurement
constraints on dark energy}, \emph{Astrophys. J.} {\bf 764} (2013) 139
[arXiv:1211.4253].

\bibitem{FarooqR13}
O. Farooq and B. Ratra, \emph{Hubble parameter measurement
constraints on the cosmological deceleration-acceleration
transition redshift}, \emph{Astrophys. J.} {\bf 766} (2013) L7,
[arXiv:1301.5243].

\bibitem{Farooqth}
 O. Farooq, \emph{ Ph.D. thesis},
arXiv:1309.3710.

\bibitem{GottV01}
J.R. Gott III, M.S. Vogeley, S. Podariu and B. Ratra, \emph{Median
statistics, h0, and the accelerating universe}, \emph{Astrophys.
J.} {\bf 549} (2001) 1.


\end{thebibliography}
\end{document}